\documentstyle[aps,prl]{revtex}

\input{epsf}

\begin{document}
 
\twocolumn[\hsize\textwidth\columnwidth\hsize\csname
@twocolumnfalse\endcsname

\title{Timing and Other Artifacts in EPR Experiments\cite{PRL/submission}}

\author{Caroline H Thompson\cite{CHT/email}\\
	Department of Computer Science, University of Wales, Aberystwyth, \\
	SY23 3DB, U.K.}

\date{\today}
\maketitle
 
\begin{abstract}
Re-evaluation of the evidence (some of it unpublished) shows that
experimenters conducting Einstein-Podolsky-Bohm (EPR) experiments may have been
deceived by various pre-conceptions  and artifacts.  False or
unproven assumptions were made regarding, in some cases, fair sampling, 
in others
timing, accidental coincidences and enhancement.  Realist possibilities,
assuming a purely wave model of light, are presented heuristically,
and suggestions given for fruitful lines of research.  Quantum
Mechanics (QM) can be proved false, but Bell tests have turned out to
be unsuitable for the task.
\end{abstract}

\vskip2pc]

Real EPR experiments are very different both from Einstein, Podolsky and
Rosen's original idea \cite{Einstein/Podolsky/Rosen35}
and from Bell's idealised situation
 \cite{Bell64}. The general 
scheme, which can have either one or two detection channels on each side,
is described in a multitude of papers \cite{Clauser/Shimony78,Selleri88}.
Bell assumed no ``missing values", or, in
other words, the certainty of detection in one channel or the other for all
values of the ``hidden variable".  As soon as real experiments were started,
it was realised that detection would {\em not} be certain, and that
consequently the inequalities would need to be modified.  The result was a
set of alternative inequalities \cite{Clauser/Horne74} (see simplified
versions, as generally used, in 
table I).

The standard inequality 
is covered in an earlier paper (``The Chaotic Ball: an Intuitive Analogy
for EPR Experiments''
\cite{Thompson96}).  Realist models that infringe it are easily
constructed if (as I consider must always be the case) there are
``variable detection probabilities''.  The current paper presents likely
explanations for (single-channel) experiments that infringe the CHSH 
or Freedman tests.

\begin{table}[!b]
\label{table1}
        \begin{tabular}{cccc}
         & Test Statistic  & Upper Limit & Auxiliary\\
        & & & Assumption\\
\hline
Standard & $S_{Std} = 4(\frac{x - y}{x + y})$ & 2 & Fair sampling\\
CHSH & $S_{C} = 3\frac{x}{Z} - \frac{y}{Z} - 2\frac{z}{Z}$ & 0 &
        No enhancement\\
Freedman & $S_F = \frac{x - y}{Z}$ & 0.25 &  "\\
\end{tabular}
\caption{Various Bell inequalities, for rotationally invariant,
factorisable
experiments.  $x = R(\pi/8)$, $y = R(3\pi/8)$,
$z = R(a, \infty)$ and $ Z = R(\infty,
\infty)$, using the usual terminology
in which $R$ is coincidence rate, $a$ is polariser setting, and
$\infty$ stands for absence of polariser.}
\end{table}

Such experiments require more than simple variable probabilities,
and very few experiments have, to my knowledge,
achieved violation.  These few are important, however, as they include the ones
that have now gained a place in the text books, namely the single-channel 
Orsay experiments \cite{Aspect81-2}.  The assumptions behind 
the CHSH and Freedman inequalities  are both the
basic one of ``factorability" (for each hidden variable value $\lambda$, the
probability of a coincidence, $p_c(\lambda)
 = p_1(\lambda).p_2(\lambda)$, where $p_1(\lambda)$ and $p_2(\lambda)$ are
the singles probabilities) and the additional one of ``no enhancement" (for
each $\lambda$ we have $p(a,\lambda) \leq p(\infty, \lambda)$).

Marshall, Santos  and Pascazio \cite{Santos85+,Pascazio89} have
suggested that the
latter assumption may be untrue, and this may well be the case
\cite{G-S/enhancement}. They have also suggested that it
could be the unjustifiable subtraction of ``accidental coincidences" in some
experiments
that causes the violations, which would mean that we have both a
kind of enhancement and a failure of factorability, but caused by erroneous
manipulation of the data rather than any fundamental property of the
experimental setup.  Though Aspect and Grangier\cite{Aspect/Grangier85}
 have countered their criticisms,
there remain severe doubts: the figures they quote
are for the two-channel experiment
\cite{Aspect82-0}, which uses the standard test.

\begin{table}
\label{table2}
\begin{tabular}{cccccccc}
 
        & $x$ & $y$ & $z$ & $Z$ & $S_{Std}$ & $S_C$ & $S_F$\\
\hline
     Raw coincidences & 86.8 & 38.3 & 126.0 & 248.2 & 1.55 & -0.121 &
0.195 \\
        Accidentals & 22.8 & 22.5 & 45.5 & 90.0 &  &  &  \\
        ``Corrected" & 64.0 & 15.8 & 80.5 & 158.2 & 2.42 & 0.096 & 0.309
\\
\end{tabular}
\caption{Effect of standard adjustment for accidental coincidences.
Data from table VII-A-1 of Aspect's thesis (1981 single-channel
experiment).}
\end{table}

My own figures 
(table II),
extracting relevant information from Aspect's thesis \cite{Aspect83-prl},
suggest that no Bell test for single-channel experiments would
have been violated had ``accidentals" not been subtracted.

The mechanism whereby accidentals can cause violation is very straightforward.
We can be fairly confident that Bell's inequalities will hold for the
raw data, but whether or not they hold after subtraction depends on
whether or not true and accidental detections are independent.  This
depends on
factors such as correlations between neighbouring emissions, how the
detector responds to close (actually overlapping, under a wave model)
 signals and instrument dead times.
The number of accidentals  (as measured by the 
number of coincidences when one stream is delayed by, say, 100 ns)
is proportional to the product of the numbers of signals on each side.
If the detectors are ``correctly''
adjusted, so that they register half the number of hits when a polariser is
inserted (which follows from Malus' Law {\em provided
noise and  settings of various voltages are appropriate} \cite{Malus_Law}),
the value is $A$, say, for terms such
as $x$ and $y$, $2A$ for terms $z$, and $4A$ for terms $Z$.  It is easily
seen that if we subtract these we increase all test statistics and hence
the likelihood of violations.

Let us review the situation, as the question of accidentals is inextricably
entangled with the whole matter of timing, choice of coincidence window
and the interpretation of time spectra.

\begin{figure}
        \centering
        \leavevmode
        \epsfbox{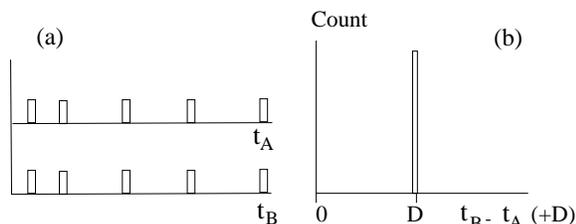}
        \caption{Assumed ideal timing and resultant spectrum.  $D$ is
the
        delay applied to the $B$ channel so as to place the peak in the
        centre of the picture.}
        \label{fig1}
\end{figure}

Proofs of Bell inequalities do not mention time.  They assume that the
source produces pairs of particles and that identification of the pairs
poses no problem.  Thus the stream of signals arriving at the coincidence
monitor (the device which in practice does the identification) can be
envisaged as in Fig.~\ref{fig1}, which also shows the expected
time-spectrum --- a single bar whose height is the number of coincidences.
Even this simple picture might have a
slight complication, as the pairs are supposed to be produced at completely
random times.  This means that there is a very slight possibility that
there will be another $B$ arrival at {\em any} interval after the $A$
(including in the same ``time-bin''), so that we expect a low constant
background of accidentals, or rather, because the instrument measures the
time to the {\em first} detection, a downward slope that is gentle provided
the emission rate is low or the probability of detection is small.

\begin{figure}
        \centering
        \leavevmode
        \epsfbox{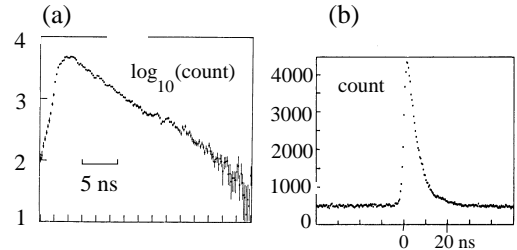}
        \caption{Actual time spectra.  (a) Freedman, (b) Aspect.
Freedman's  could have been slightly distorted by the
instrumentation.  It is drawn by hand and represents the combined
results of a whole series
of runs.  Aspect's is for just one run and would have been displayed on
a
VDU (actual runs would have had considerably greater scatter as they
were
over shorter periods)}
        \label{fig2}
\end{figure}

Consider now some spectra (histograms of differences in detection times
of $A$ and $B$ signals) from actual experiments, Freedman's of 1972
\cite{Freedman72}
and Aspect's of around 1980 (Fig. \ref{fig2}).
 Numbers of coincidences can be estimated by defining
an ``integration window'' and organising electronics so as to count all signals
that arrive within it.  It is conventionally described in terms of just the one
parameter, $w$, though in reality it requires two --- a start time relative
to the peak of the spectrum, as well as
the window length.  Aspect in his final experiment used a window from $-3$~ns
to $+17$~ns.  Freedman used one of just 8~ns length, but does not tell
us how the start was chosen.  Note that the difference in shape between
the two spectra is due primarily to the scales --- log in one case, not
in the other.

From their PhD theses, it is clear how the experimenters were interpreting their
time-spectra.  Both Aspect and Freedman were thinking of the decreasing
region as showing the distribution of emission times (controlled by 
the ``lifetime of the intermediate stage of the cascade'')
of the second
``photon", this being regarded as a particle.  They took the rising front
as due to
{\em random}, normally distributed, timing variations.  
There was a constant underlying background
of accidentals.

  Now under a classical wave theory of
light, this interpretation is not reasonable on several counts.  
Firstly, various
considerations (see above) make the assumption
of constant accidentals implausible.  
Secondly, if the rise were due just to
random variations, then would not the peaks of the spectra be broader?
(Interestingly, Aspect mentions that the variations do cause a broadening, but
this is not evident here.) Under wave theory, the results seem to
indicate rather that the $A$ and $B$ ``photons" are emitted simultaneously,
and each is a wave that starts at high intensity and decreases at roughly
negative exponential rate, only the $A$ one much more fast than the $B$.
(We assume that detection occurs when the addition of 
electromagnetic noise to the
signal pushes the total over some threshold.)
This interpretation becomes more obvious if one thinks of the pairs of {\em
equal} ``photons" of the Stirling experiments \cite{Perrie85}, in which
the time-spectra are symmetrical.  There is possible experimental support for
this view: it might explain Aspect's problems with ``{\em post-impulsions}'', 
or multiple detections, some of which occurred despite dead times of 16~ns
or more.

If we admit, however, that the rising front is not solely the result of
random time variations --- and they are, in any case, most unlikely to
be normally distributed --- this raises the question of how we interpret the
various observed timing variations.
They are only small (standard errors of about 0.7 ns for each
photomultiplier and 0.1 ns for each discriminator, according to Aspect's
investigations \cite{Aspect83-prl}, but if they were systematically related to
the polarisation angles --- which they would be if related to {\em
intensity} of the input \cite{Timing/intensity} --- then they could
contribute to a systematic change
in the quality of synchronisation between different values of $\phi$, the
angle between polariser settings.
This could, if windows were too small to
include all genuine coincidences, mean that factorability did not hold
exactly \cite{Fine82+}.  It could mean that 
$R(\pi/8)$, involving the smaller angle, was based on pairs that were
slightly better synchronised than $R(3\pi/8)$, giving a slight increase to
the difference and hence a slight increase in the chance of violating a
Bell inequality.  

Aspect was aware of the need to ensure that his
window included all true coincidences.  His theoretical calculations
suggested that he included 97\% of them, but he never mentions the 
danger of systematic time variations, which could mean slight
differences in shape between spectra for different $\phi$.  He would not
have been able to detect such differences by eye, due to scatter 
and accidentals, and it is impossible to tell whether
or not his chosen window was adequate.  Freedman's was too small unless
accepted theory is absolutely perfect.  Both Aspect and Freedman used as
their main criterion the  maximisation of a ``quality factor''.  This
amounted to minimising the running time of the experiment --- surely of
lesser importance than the avoidance of systematic error?

A slight timing effect may well be present in all
cascade experiments, but is unlikely to be the prime cause of inequality
violations.  As mentioned above, it is more likely in Aspect's experiments
that the accidental coincidence subtraction is the prime cause, and if we
think again about Freedman's spectrum (Fig.~\ref{fig2} (a)), we might
suspect from what he tells us of the conditions under
which it was obtained that there is a rather more crude effect coming into
play here.  It could be simply a matter of presence or absence of the
polarisers producing small changes in transit times, shifting the whole
spectrum.  The polarisers were very large (``pile of plates'' type,
180~cm long), so, though the 
thickness of the glass in them was small, if there were multiple reflections
 a nanosecond or so of variation might occur.  Poor choice of start and 
end of his very short window could then easily cause
bias towards lower values with polariser absent, which could lead to
inequality violation.  Aspect, it should be noted, ``corrected" for any such
effect by adjusting a variable delay.

This has brought us a long way from Bell's simple idea, and the limits of
complexity of the real situations have only just started to be explored.
Much more could be said about the effect of our assumptions about the
emission process (Is it really random, or is there any slight tendency
to clustering, or, conversely, to even spacing in time?);
 about the effects of dead times (various
different ones come into play at different points in the experiments,
having the effect of (a) suppressing multiple detections and (b) adding yet
one more difference between the sets of signals that constitute ``singles"
and the ``coincidences");  and about the detection process (Is it necessarily
``square law" for the full range of signals encountered?  Is the electromagnetic
noise that is an essential part of the process \cite{Gilbert/Sulcs96}
steady or does it fluctuate?). 

 My own interest
has been in exploring the possibility that time spectra may have extended
tails when produced by attenuated signals.  These tails would
arise if the signals were in fact decaying only very slowly and the
probability of detection per unit time were high. (It would be
sufficient to have the probability high for some subset of the whole set
of signals.)  The effect arises because
the time spectrum shows the time to the {\em first detection}
only, so that the probability for a given time-delay may be greater if the
probability that the current signal was {\em not} detected prior to this is
smaller.  The fascination of this subject was the possibility of a completely
different demonstration of the pure wave nature of light, in the time-domain
instead of the  familiar spatial domain of the two-slit experiment.  It is
interesting that experimenters  assume that measuring an atomic lifetime using
time-spectra (which measure the  {\em first} detection times) is exactly
equivalent to a method in which many ``photons'' are detected simultaneously
and produce directly a time-varying electric signal.  A paper detailing the
method is quoted by Aspect as an authority on measurement of lifetimes
\cite{Havey77}. Computer simulations have shown that
experiments to date have probably not been suitable for
showing this effect, but it might be possible to devise one, with very
low emission rate or with the more controllable pairs of signals
produced in parametric down-conversion.

\begin{figure}
        \centering
        \leavevmode
        \epsfbox{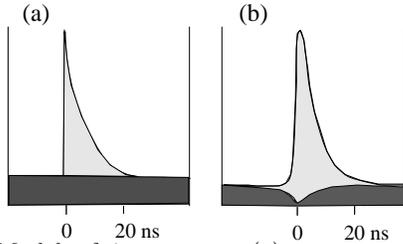}
        \caption{Models of time spectra: (a) quantum mechanics
assumption
and (b) conjectural realistic model.  Light shading: ``true''; Dark
shading: ``accidental'' coincidences.}
        \label{fig3}
\end{figure}

To return to matters of more immediate importance, Aspect's experiments
involved  seriously large accidental coincidences.  As he says, there could
typically be 600 accidentals to 200 true coincidences displayed on the VDU.  
One can seriously
question, therefore, whether it is possible to extract a valid Bell-type
test from such data.  His idealisation is illustrated by Fig.~\ref{fig3}(a),
taken  from his thesis.  But we have no independent way of judging the
true picture.   This (if one can be said to exist) might be
as in Fig.~\ref{fig3}(b).

Freedman's experiment is perhaps the only one to have been performed in
which accidentals do not appear to be important --- we get violation
of inequalities even when they are not subtracted --- but, as explained
above, it used a coincidence window that was too small.  It was wide
open to synchronisation problems.
 
There appears to have
 been a serious miscarriage of science in accepting existing
experiments as supporting quantum theory.  Logically, this hypothesis
should have been rejected at the outset, as the implied non-local
effects are impossible.  It is evident that existing classical
explanations are also  wrong, as they give wrong predictions.  It should
therefore have been realised that {\em all} existing theory should have
been challenged.  It should not have been possible for Aspect to make
statements such as (translating loosely from his thesis): ``Agreement
with quantum theory is a privileged method for confirming that the
apparatus is correctly set'' (referring, presumably, not to the final
conclusion of Bell violations but to intermediate decisions such as
conformity to Malus' Law).  Freedman concluded his thesis with a remark
to the effect that there was no need to search too hard for causes of
systematic error as Bell's inequalities had been violated and were of
such general applicability.  Many workers have allowed themselves to be
influenced by an opinion that any imperfections would bring quantum
theory predictions nearer to classical ones.  Perversely, this has
turned out to be true, but it is the classical ones that are brought
nearer to  quantum theory, not the other way around!

To conclude, I should like to take this opportunity to state some general
opinions.  Firstly, the setup of EPR experiments could be put to much
more constructive use than merely attempting violate Bell inequalities.
By exploring objectively a wider range of parameters, with detectors
purposely set ``wrongly" (so that we do not get Malus' Law reproduced or
neat results such as doubling coincidences when we remove a polariser),
these experiments could {\em prove} that light is purely wave --- that it
is not a matter of the polariser passing a certain percentage of ``photons"
but of it reducing the intensity of each signal, in exactly the same way as
on a classical level though (see Marshall and Santos's Stochastic Optics work) 
being subject,
due to the very low intensities, to random variations from additions of
background electromagnetic noise.  Secondly, I would suggest that computer
simulations should be conducted in parallel with the experiments. 
The act of constructing the computer model brings home the logical structure,
which cannot possibly be that of the Quantum Mechanics collapsing
wavefunction.  This, as Feynman showed \cite{Feynman82}, cannot be simulated.

\bibliographystyle{prsty}

\begin{references}
\bibitem{Malus_Law}
Malus' Law should not be assumed in EPR
experiments.  It may well be that it holds exactly for {\em
intensities}, but a realist theory of detection must admit the
likelihood
that the probability of detection is {\em not} exactly proportional to
the intensity.  Hence Malus' Law would not in general hold exactly for
the measured counts.
\end{references}

 
\end{document}